\documentstyle[preprint,epsfig,aps]{revtex}
\newcommand{\mbf}[1]{\mbox{\boldmath$#1$}}

\begin{document}

\title { Configuration Mixing in Relativistic Quark Models}
\author{H. J. Weber }
\address{ Institute of Nuclear and Particle Physics,
University of Virginia, \\ Charlottesville, VA 
22903, USA }
\author{M. Beyer}
\address{Fachbereich Physik, Universit\"at Rostock, 18051 Rostock, Germany}
\maketitle
\begin{abstract}
Starting from the standard and commonly used relativistic nucleon wave
function, we show that residual interactions based on gluon or
Goldstone boson exchanges generate additional components leading to 
configuration mixing.  This result suggests that realistic nucleon
wave functions can be expected to be far more complex than seemingly
successful form factor fits in relativistic quark models
would have one believe.
\end{abstract}
\vskip0.5in
\par
PACS numbers: 12.39.Ki,\ 14.20.Dh
\par
Keywords: Nucleon, configuration mixing, light cone quark model 
\newline

\section{Introduction}

The nonrelativistic quark model (NQM) explains qualitatively many of
the strong, electromagnetic and weak interaction properties of the
nucleon and other octet and decuplet baryons in terms of three valence
quarks whose dynamics is motivated by quantum chromodynamics (QCD),
the gauge field theory of the strong interaction. Lacking presently a
complete solution of QCD at low and intermediate energies, where
nonperturbative effects are essential, quark models are useful as they
provide physical insight and can easily be related to empirical
analyses. However nonrelativistic approaches face difficulties, when
quantitative results are required, e.g., at larger momentum transfers,
or even at $q^2=0$ for some weak interaction observables\footnote{One
of the early and obvious examples for a relativistic effect is the
parity violating weak current that leads to the axial coupling of the
nucleon to the $W$ boson, $g_A$~\cite{mitbag}.}.

A frame for a systematic approach to a relativistic few-body theory is
provided by light front dynamics~\cite{dir47}. This is intuitively
appealing since it uses a Hamiltonian concept and is thus easy to
interprete. As in the nonrelativistic case the phenomenon of
confinement is modeled by a confining force that can be chosen to be
three dimensional. Residual interactions between quarks related to
``hyperfine'' splittings may be calculated in light cone time ordered
perturbation theory. By now several residual interactions have been
used in the context of nonrelativistic models, gluon type, Goldstone
boson type and instanton type~\cite{thofft} interactions that lead to
characteristic spin, isospin, and orbital mass splittings in the
baryonic spectrum~\cite{KI,bonn,plessas}. All these interactions
induce configuration mixing that beside mass splitting shows
qualitatively new effects. None of these interactions have been
considered in the light front framework so far and merely ``ground
state'' wave functions of the nucleon have been used. Since a complete
solution of the relativistic three-quark problem on the light front is
still missing, we provide a further step towards this aim and
incorporate the Goldstone boson and the gluon exchange: We calculate
the spin-isospin matrix elements that appear in the three body
equation. We show that both interactions lead to configuration
mixing. 

Due to the spontaneous chiral symmetry breakdown ($\chi $SB) of QCD,
the effective degrees of freedom at the scale $\Lambda _{QCD}$ are
quarks along with Goldstone bosons. Chiral field theory involves the effective 
strong interaction commonly used in chiral perturbation theory 
($\chi$PT~\cite{CPT}) and
applies at scales from $\Lambda _{QCD}$ up to the chiral symmetry
restoration scale $\sim\Lambda _{\chi }=4\pi f_{\pi }\sim 1.17$ GeV,
where $f_{\pi }=0.093$ GeV is the pion decay constant. It provides a
scenario inside hadrons at distances that are smaller than the
confinement scale $\Lambda ^{-1}_{QCD}$, but larger than distances
where perturbative QCD applies, in which the effective degrees of
freedom of the strong interaction are dynamical quarks (with a mass
depending on their momentum) along with the SU(3) octet of Goldstone
bosons comprising the pions, kaons and the eta meson.  Chiral field
theory dissolves a dynamical quark into a current quark
and a cloud of virtual Goldstone bosons.

\section{Relativistic Nucleon Wave Function  }

Within the nonrelativistic quark model (NQM) baryons are described as
Fock states of three quarks - in the simplest case. Although not
derived from first principles of QCD, the bulk properties of the
nucleon can be described in a qualitative fashion. Meanwhile 
experience in quark modeling and various experiments at
high energies established the necessity of using a relativistic
description of the nucleon. The
difficulties of a proper relativistic description of bound states are also 
obvious and, to date, neither a general solution exists nor one
that would satisfy all required conditions. Among the promising
candidates for a covariant description of a few-body bound state is
one utilizing the light front form. In the context of the two-body
problem the connection between the four dimensional covariant form of
the Bethe-Salpeter amplitude and the three-dimensional light front
wave function has been described explicitly and their differences have
been discussed in Ref.~\cite{bon98}.

So far the three quark bound state problem on the light front for the
nucleon has been tackled utilizing several (sometimes drastic)
approximations, e.g. spinless quarks within a Weinberg-type
or a Faddeev-type equation~\cite{sal95,TF}. To evaluate
currents with appropriate boost properties the wave functions can be
constructed by rewriting the nonrelativistic results in light front
variables using Melosh rotated Pauli spinors~\cite{KP}.

A particular problem in the three quark system is the proper modeling
of the confining force that presently seems to require a three
dimensional form.

Lacking at present a complete solution our strategy is as follows. For
the time being we assume that the three quark wave function can be
expanded into a set of square integrable functions. As an example
harmonic oscillator functions have widely been used in the past.
Indeed one may assume a three dimensional harmonic oscillator
confining potential $U$ that will be added to the mass (squared)
operator, $M_0^2$, of the free three quark system following the
Bakamjian-Thomas construction. This leads to harmonic oscillator
functions $\{R_n\}$ for the configuration/momentum space. For the
spinless case this approach has been shown to lead to the proper Regge
trajectories, both for mesons and baryons, provided some conditions on
the parameters are fulfilled~\cite{sal95}.

The proper coordinates to describe the evolution of a particle in
light cone dynamics are the light cone time $\tau=x^0+x^3$ and the
coordinates $(x^-,\mbf{x}_\perp) = (x^0-x^3,(x_1,x_2))$ arranged as
$x=(\tau,x^-,\mbf{x}_\perp)$. The momenta of the quark in the nucleon moving 
frame are denoted by $p_i$ and
\begin{equation}
p^\mu = \left(p^+=p^0+p^3,p^-=\frac{m^2+\mbf{p}^2_\perp}{p^+},
\mbf{p}_\perp\right).  
\end{equation}
In the light-cone approach all particles are on their mass shell, i.e.
$m^2=p^+p^--\mbf{p}^2_\perp$ and the spectrum of the total momentum,
$P^+$, is bounded from below. For all particles $p_{i}^{+}\ge 0$, and the
kinematically invariant light cone fractions are $x_i=p_i^+/P^+$.
Since light cone quarks are off the $p^-$ shell
the three-body proper coordinates are given in terms of the $+$ and
$\perp$ components only,
\begin{eqnarray}
P&=&\sum_i p_i,\nonumber\\
q_3 &=& \frac{x_2p_1-x_1p_2}{x_1+x_2},\nonumber\\
Q_3 &=& (1-x_3) p_3 - x_3 (p_1+p_2),
\end{eqnarray} 
where the first equation is the total momentum $P$ and the other two
are coordinates for the pair $q_3$ and for the odd quark $Q_3$ (given
in the (12)3 coupling scheme, viz. channel 3. Other channels are given
by cyclic permutation of 1,2,3.). Note that $q_3^+=Q_3^+=0$.

Several boosts are kinematic, whereas rotations ($\mbf{L}_{\perp}$) are not. 
The kinematic light cone generators are transitive on the null plane, 
so that the wave function is known everywhere, once it is known in the
rest frame of a bound few-body system. Finally, the total momentum
separates rigorously from the internal motion in momentum space, which is not 
the case in the instant form, where all boosts are interaction dependent and 
rotations are kinematic.    

By now the structure of the ground state (i.e. without additional
momentum excitations and with symmetric momentum space wave function) in the
light front formalism is well understood and tested. In the rest frame of the
three particle system with free total momentum, 
$\stackrel {\circ}{P} =(M_0,M_0,{\bf 0}_\perp)$, the nucleon ground state  
may be written in the following way:
\begin{equation}
\Psi _N(123)=\psi _N(123)\
R_0(\mbf{k}_{i,\perp},k_i^+),\ 
\label{wf1}
\end{equation}
where $1\equiv \{{k_1},\lambda _1,\tau _1,\dots\}$ denotes the
momentum $k_1$, helicity $\lambda _1$, isospin $\tau _1$ of particle 1
and $\psi _N$ is the relativistic wave function depending on spin and
isospin. The Dirac structures necessary to specify the spin-isospin
components are given in Table~\ref{tab:1}. However, because of the
symmetry restrictions ($S_3$ and parity of the total wave function)
the number of independent ground state components reduces to three
linearly independent symmetric spin-flavor states (see
e.g. \cite{WAP} and refs. therein), viz.~\footnote{Different
momentum wave functions, $R_{0}^{(\kappa )}$, for various
spin-flavor invariants $I_{\kappa }$ are not ruled out and would, in
fact, be needed to describe different momentum dependence of the
proton's charge and magnetic form factors according to recent
measurements at the Thomas Jefferson Research Laboratory (JLab)~\cite{CP}}
\begin{eqnarray} 
\psi _N(123) = \sum_{\kappa =0}^2 c_\kappa I_{\kappa }(123)\ u_1\,u_2\,u_3.
\label{wf2}
\end{eqnarray}
The quark spinors are denoted by $u_i=u_{LC}(p_i,\lambda _i)\chi_i$
where the light cone spinors $ u_{LC}(p_i,\lambda _i)$ are given explicitly,
e.g., in Ref.~\cite{kon90} and $\chi_i$ are the isospinors. The
spin-isospin wave functions invariants are given by
\begin{eqnarray}
I_0(123)&=&\bar{u}_{1}\gamma _5 G\bar{u}_{2}^{T}\
\bar{u}_{3}u_{\lambda }+(23)1+(31)2,\\
I_1(123)&=&\bar{u}_{1}\gamma ^{\mu}
\mbf{\tau }G\bar{u}_{2}^{T}\
\bar{u}_{3}\gamma _5\gamma _\mu
\mbf{\tau }u_{\lambda }+(23)1+(31)2,\\
I_2(123)&=&\bar{u}_{1}\gamma \cdot P \gamma _{5}G\bar{u}_{2}^{T}\
\bar{u}_{3}u_{\lambda }+(23)1+(31)2,
\label{eqn:I012}
\end{eqnarray}
where $G=i\tau _2 C$, and $C=i\gamma _0\gamma _2$ is the charge
conjugation matrix and $T$ denotes transposition. The coefficients
$c_{\kappa }$ have to be determined from a dynamical equation. The
larger number of nucleonic components even in the ground state
(compared to only one in the nonrelativistic case) is related to
negative energy contributions in the wave function. This is similar to
the deuteron as a bound two nucleon system and gives rise to $P$
states that are of pure relativistic origin. In the past, the
coefficients $c_{\kappa }$ have been chosen in a way that the
resulting ground state relativistic wave function is restricted to
contain only positive energy components. As a consequence only one
independent invariant spin-flavor structure arises. This can best been
seen in the Bargmann Wigner basis~\cite{BKW}, but is also clear from
the nonrelativistic wave function that is Melosh boosted to the light
cone. In its turn this particular relativistic wave function can be
said to be a straightforward generalization (based on Melosh
rotations) of the nonrelativistic three quark $s$-wave function (that
contains positive energy Fock components only). This particular linear
combination, $G_2+G_6$, is then given by
\begin{equation}
\psi_{0}(123)={\cal N}\sum_{\lambda_i} \left[\left(\bar{u}_{1}(\gamma \cdot
P+M_{0}) \gamma _{5}G\bar{u}_{2}^{T}\right)
\left(\bar{u}_{3}u_N(P)\right)+(23)1+(31)2\right]u_1u_2u_3,\
\label{eqn:psiN}
\end{equation}
where $\cal N$ denotes the normalization and the nucleon spinor
$u_N(P)$ depends on total $P$. In other words, when the mixed
antisymmetric nucleon spin-isospin state is Melosh boosted, the
combination $G_2+G_6$+permutations of Table~\ref{tab:1} and
Eq.~\ref{eqn:psiN} is obtained.  When the mixed symmetric state is
similarly treated another combination, $G_3-G_5-G_8$, is obtained. Of
course, in the $uds$ basis both lead to the same totally symmetric
spin-isospin nucleon wave function. The excited states may be
constructed in a similar fashion, as has been shown in
Ref.~\cite{BKW}.

\begin{figure}
\leavevmode
\begin{center}
\psfig{figure=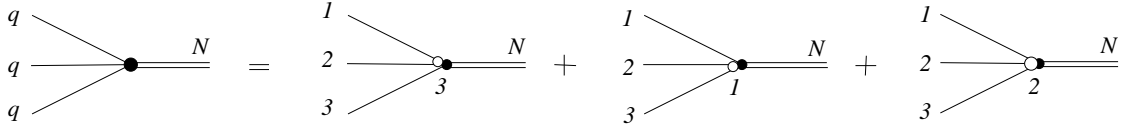,width=0.9\textwidth}
\end{center}
\caption{\label{fig:wfkt} Faddeev components of the wave function, see
Eq.~\protect\ref{eqn:psiN}.}
\end{figure}

\begin{figure}
\leavevmode
\begin{center}
\psfig{figure=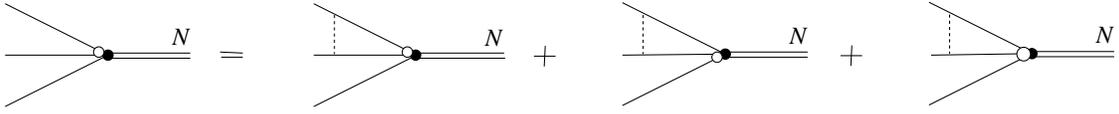,width=0.9\textwidth}
\end{center}
\caption{\label{fig:eqn} Pictorial demonstration of
Eq. \protect\ref{eqn:fad} for $\Psi_N^{(3)}$.}
\end{figure}

The equation of motion for the three quark bound state $\Psi_N$ may be
written~\cite{sal95,wei66,fra81} according to the Bakamjian-Thomas 
prescription (BT) as
\begin{equation}
\Psi_N = G_3^{(0)}(m_N^2)  (V^{(1)}+V^{(2)}+V^{(3)}) \Psi_N,
\label{eqn:psi}
\end{equation}
where we have assumed a two-body interaction $V$ and utilized the three-body
channel notation (i.e. $V^{(3)}=V(12)$ etc.). The free Green function
$G_3^{(0)}$ is given by
\begin{eqnarray}
G_3^{(0)-1}(m_N^2) &=& P^+\left(P^--\sum_{j=1}^3p_j^-\right)
\nonumber\\
&=&m_N^2 + \frac{1-x_3}{x_1x_2}\,q_3^2+\frac{1}{x_3(1-x_3)}\,Q_3^2
-\sum_{j=1}^3\frac{m_j^2}{x_j}
\end{eqnarray}
that has to be evaluated at the eigenvalue $m_N$ of $\Psi_N$, and
$q_3$ and $Q_3$ are four-vectors. Note that $G_3^{(0)}$ is independent
of the specific channel chosen for the three body coordinates (here
channel 3). The $uds$ basis used in Eqs.~\ref{eqn:I012} and \ref{eqn:psiN} 
suggests a Faddeev decomposition of the wave function,
$\Psi_N=\Psi^{(1)}_N+\Psi^{(2)}_N +\Psi^{(3)}$ shown in
Fig.~\ref{fig:wfkt}.  In this notation the Faddeev components of the
three-body bound state are given by ($\alpha =1,2,3$)
\begin{equation}
\Psi_N^{(\alpha)} = G_3^{(0)} (m_N^2)\ V^{(\alpha)}\,  
\Psi_N,\qquad\alpha=1,2,3. 
\label{eqn:fad}
\end{equation}
A diagrammatical form of this equation is shown in Fig.~\ref{fig:eqn}.
The integrals appearing in Eqs.~\ref{eqn:psi} and \ref{eqn:fad} are
written more explicitely (e.g. in channel 3)
\begin{eqnarray}
(V^{(3)}\Psi_N)(1'2'3')  &=& \sum_{123}\
V^{(3)}(1'2',12)\ \delta_{3'3}
\Psi_N(123)\nonumber\\
&=& \int \frac{d^2\mbf{q}_{3\perp}dx_1}
{2(2\pi)^3x_1}\
V^{(3)}(x'_1,\mbf{q}_{3\perp}';x_1,\mbf{q}_{3\perp})\
\Psi_N(123'),
\label{eqn:int}
\end{eqnarray}
The last equation is written for $\mbf{P}_\perp=0$, and spin isospin
indices suppressed. Note that $x_1+x_2+x_3'=1$ must be fulfilled.

\begin{figure}
\leavevmode
\begin{center}
\psfig{figure=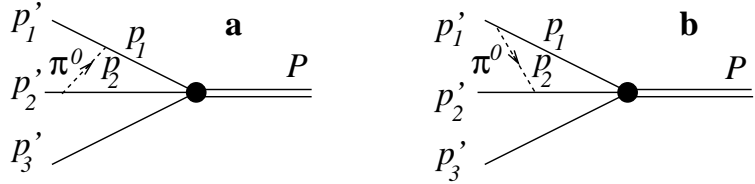,width=0.6\textwidth}
\end{center}
\caption{\label{fig:pi0} Light cone time ordered neutral pion exchange
between quarks 1 and 2. }
\end{figure}

\begin{figure}
\leavevmode
\begin{center}
\psfig{figure=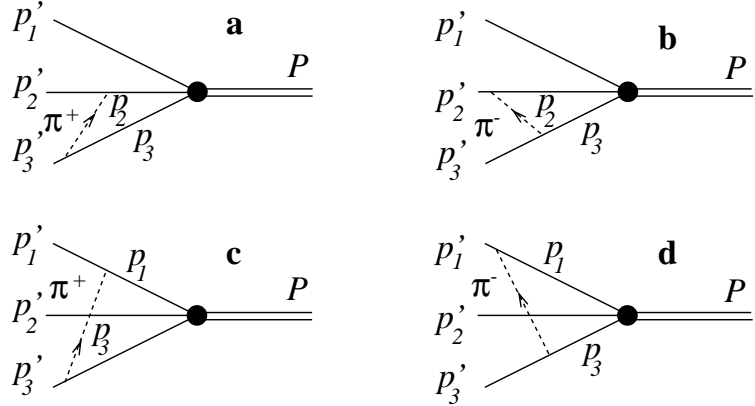,width=0.6\textwidth}
\end{center}
\caption{\label{fig:pi} Light cone time ordered charged pion
exchange.}
\end{figure}

\begin{figure}
\leavevmode
\begin{center}
\psfig{figure=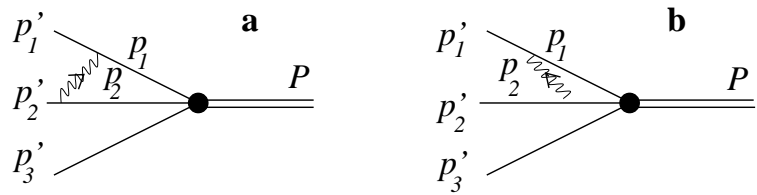,width=0.6\textwidth}
\end{center}
\caption{\label{fig:gl}Gluon exchange between quarks 1 and 2. }
\end{figure}

This integral will now be evaluated in the following way: We introduce
the nucleon wave function given in Eq.~\ref{wf1} and calculate the
corresponding diagrams of Figs.~\ref{fig:pi0}-\ref{fig:gl} using light
front perturbation theory.

The diagram of Fig.~\ref{fig:pi0}a, e.g., represents the $\pi^0$
exchange between quark 1 and 2 with momentum $k=p_1-p_1'$ for the
$\perp,+$ components. In this case the integral, Eq.~\ref{eqn:int}
becomes  (up to a Jacobian and constants),
\begin{equation}
 \int \frac{dk^+ d^2\mbf{k}_\perp}{2(2\pi)^3k^+} \ 
\frac{v^{(3)}(k^+,\mbf{k}_\perp)}{E_{12}}\
 R_0(k^+,\mbf{k}_{\perp},3')\,\psi_N(123').
\label{eqn:iter}
\end{equation} 
The light front energy denominator
$E_{12}$ for this particular case is given by
\begin{equation}
E_{12}=m_N^2-E_3'-E_\pi-E_1'-E_2,
\label{eqn:den}
\end{equation}
with on-shell light-cone energies $E_i$ and
$E_{\pi}$, while the remaining
vertex function denoted here by $v^{(3)}=v(12)$ will be defined in the
next section.

In the remaining part of the paper we investigate this integral and
evaluate the spin isospin terms. Presently we focus on two types
of residual interactions between the quarks, Goldstone boson exchange
and gluon exchange. We will show that both interactions lead to
configuration mixing in the ground state (nucleon) wave function that
has not yet been considered for relativistic quark models.

We note here that for the two nucleon case integrals of this type have
been evaluated for the Bethe-Salpeter approach~\cite{bon98} and the
light front approach~\cite{car98}. The respective equations can be
solved by iterations. Starting from a nonrelativistic wave function
already the first order approximation generates all the relativistic
components and has been quite useful to study qualitative features of
the wave functions.  The additional wave function components arising
are indeed essential for the description of deuteron break up even at
threshold energies as they are related to the standard relativistic
pair current corrections in a nonrelativistic frame work,
see~\cite{bon98}.  In the case of magnetic moments of the deuteron
these additional wave function components have been shown to
quantitatively agree with the standard expansion in terms of pair
currents~\cite{bon99}.

\section{Goldstone Boson and Gluon Exchange}

We discuss two different types of residual interactions. One based on
gluon exchange ($v_G$) has been thoroughly studied in nonrelativistic
models. Another based on Goldstone boson exchange ($v_5$) has recently
been suggested as a possibility to incorporate the concept of chiral
symmetry breaking into quark modeling~\cite{plessas}.

If the chiral symmetry breakdown is based on $SU(3)_L \times SU(3)_R$,
then the effective interaction between quarks and the octet of
Goldstone boson (GB) fields $\Phi_i$ involves the axial vector coupling    
\begin{eqnarray}
{\cal L}_{int}=-{g_A\over 2f_\pi }\sum_{i=1}^{8}\bar q \partial_\mu 
               \gamma ^\mu \gamma _5 \lambda _i \Phi_i q             
\label{lint}
\end{eqnarray}
that is well known from soft-pion physics. In Eq.~\ref{lint}, the 
$\lambda _i$, $(i=1,2,...,8)$ are Gell-Mann's SU(3) flavor 
matrices, and $g_A$ is the dimensionless axial vector-quark coupling constant 
that is taken to be one here. As a consequence, the polarization of quarks 
flips in chiral fluctuations, $q_{\uparrow,\downarrow,}\rightarrow 
q_{\downarrow,\uparrow}+GB$, into pseudoscalar mesons of the SU(3)
flavor octet of Goldstone bosons, but for massive quarks the
non-spinflip transitions from $\gamma_{\pm}\gamma_5 k^{\pm}$ that
depend on the quark masses are not
negligible. Let us also emphasize that, despite the nonperturbative 
nature of the chiral symmetry breakdown, the interaction between quarks and 
Goldstone bosons is small enough for a {\em perturbative expansion} to apply. 

The nucleon wave function based on the Pauli-Melosh basis contains no
small Dirac spinor components. Hence its spin function is uniquely
specified by the S-wave 3-quark wave function in the case of no
configuration mixing. In terms of the Dirac-Melosh basis its spin wave
function involves the combination $M_0 G_2 + G_6$~\cite{BKW} of spin
invariants. Many electroweak form factor calculations are based on
this wave function. We wish to show first that effective interactions
like pion exchange or gluon exchange generate many other relativistic
spin invariants starting from this component alone.  

To be specific we will drop the momentum wave function of the
proton and evaluate the term $v_5(12)\psi_N$ that is the spin-isospin
part of the ground state (nucleon) wave function.  That
is, we consider a 1-2 symmetric neutral pion exchange interaction
in channel 3 of the nucleon equation of motion (see Eq.~\ref{eqn:fad},
Figs.~\ref{fig:eqn} and \ref{fig:pi0}, $\pi ^0$ exch. with $uud$
flavor part), i.e.
\begin{eqnarray}\nonumber
v_5(12)\psi_N
&=&v_5(12)(\psi^{(1)}_N+\psi^{(2)}_N+\psi^{(3)}_N)
\\\nonumber
&=&4m_q^2\ \bar u'_1 \gamma _5 u_1(\bar u'_2 
\gamma _5 u_2)\left(\bar u_3 [P] \gamma _5 C\bar u^{T}_{1} (\bar u_2 u_N)
+\bar u_3 [P] \gamma _5 C \bar u^{T}_{2}(\bar u_1 u_N)\right)\\
&=& -\bar u'_3[P]\gamma \cdot \bar k C \bar u'^{T}_{1} 
 (\bar u'_2 \gamma _5\gamma \cdot k u_N)
  -\bar u'_3[P]\gamma \cdot \bar k C\bar u'^{T}_{2}
 (\bar u'_1 \gamma _5\gamma \cdot \bar k u_N).
\label{me1}
\end{eqnarray}
The positive-energy Dirac projection operator $[P]\equiv (\gamma \cdot
P+M_0)$ in Eq.~\ref{me1} originates from the Melosh boost of the
quarks to the light cone as explained in the previous section. Here,
the isospin part is treated separately and therefore the
light cone quark spinors are given by $u_i=u(p_i,\lambda _i)$,
$u'_i=u(p'_i,\lambda '_i)$. Note that
the second term in Eq.~\ref{me1} is related to the first by
$1\leftrightarrow 2$ symmetry.
 
The instantaneous quark propagator has been included in the quark propagator 
(i.e. spin sum) by replacing the quark minus momentum component $p^{-}_{i}$ 
by the modified value~\cite{JN} 
$\tilde{p}^{-}_{i}=P^--\bar p^{-}_{j}-\bar p^{-}_{k}-
\bar k^{-}$, where the 
bar denotes the on-shell quantity and $k$ the pion momentum. For $v_5(12)$, 
$\tilde p^{-}_{2}$ for example, may be written as 
\begin{eqnarray}\nonumber
\tilde p^{-}_{2}&=&(P^{-}-\sum_i \bar p^{-}_{i})
+\bar p^{-}_{2}-\bar k^-
\\
&=&\Delta P+(\bar p^{-}_{2}-\bar p'^{-}_{2}-2\bar k^-)+\bar p'^{-}_{2}
+\bar k^- \equiv \Delta p_2+\bar p'^{-}_{2}+\bar k^-,\\
\Delta P&=&P^--\sum_{i=1}^{3}\bar p^{-}_{i}. 
\end{eqnarray}  
Hence, when we sum 
over intermediate quark helicities for the first term in Eq.~\ref{me1} using  
the kinematics $p'_1=p_1-k$, $p'_2=p_2+k$ for the $+, \perp$ components in 
$\sum_{\lambda _2} u_2\bar u_2=(\gamma \cdot p_2+m_q)/2m_q$ in conjunction 
with the free Dirac equation for $u'_2$, we obtain 
\begin{eqnarray}\nonumber 
2m_q\sum_{\lambda _2} \bar u'_2\gamma _5u_2\ \bar u_2 u_N
&=&\bar u'_2\gamma _5(\gamma \cdot \tilde p_2+m_q)u_N\\
&=&\bar u'_2\gamma _5\gamma \cdot \bar k u_N+\Delta p_2(\bar u'_2\gamma _5
\gamma ^+u_N). 
\label{offsh}
\end{eqnarray} 
The small off-minus-shell second term in Eq.~\ref{offsh} (and elsewhere) will 
be neglected in the following for simplicity and merely indicated by $\cdots$, 
but will be considered and numerically investigated elsewhere. Note that    
the pion momentum $k$ is the integration variable in the nucleon 
equation of motion. 

In order to generate nucleon spin invariants in Table ~\ref{tab:1}, we
Fierz transform each term on the right of Eq.~\ref{me1} using the
scalar line of Table~\ref{tab:2}.  This produces five
terms which collapse to the following two upon using their
($1\leftrightarrow 2$) symmetry or antisymmetry,
\begin{eqnarray}\nonumber
v_5(12){\psi }_N &=& -\frac{1}{2}\ \bar u'_1 \gamma _{
\mu } C \bar u'^{T}_{2} \ (\bar u'_3[P]\gamma _5\gamma \cdot k \gamma ^{\mu }
\gamma \cdot k u_N)\\
&&+\frac{1}{4}\;\bar u'_1 \sigma _{\mu \nu } C \bar u'^{T}_{2}\
(\bar u'_3 [P] \gamma _5 \gamma \cdot k  \sigma ^{\mu \nu }\gamma \cdot k u_N),
\label{me2}
\end{eqnarray}
whose 1-2 symmetric first matrix elements already have the canonical form of 
the nucleon spin invariants of Table~\ref{tab:1}. The second part is 
reorganized to yield 
\begin{eqnarray}\nonumber
v_5(12)\psi_N &=& -\bar u'_1 \gamma _{\mu } 
C \bar u'^{T}_{2}\ \bar u'_3\gamma _5 \Big[-M_0 k^2\gamma ^{\mu } +k^2 P^{\mu }
+2M_0 k^{\mu }\gamma \cdot k-2k\cdot P k^{\mu }\Big]u_N\\ \nonumber
&& +{1\over 2}\bar u'_1 \sigma _{\mu \nu} C \bar u'^{T}_{2}\; \bar u'_3 
\gamma _5 \Big[i k^2(\gamma ^{\mu }P^{\nu }-\gamma ^{\nu }P^{\mu })+2i(P^{\mu }
k^{\nu }-P^{\nu }k^{\mu })\gamma \cdot k\\ 
&& \qquad\qquad\qquad\qquad\qquad
+2ik\cdot P(k^{\mu }\gamma ^{\nu }-k^{\nu }\gamma ^{\mu })\Big] u_N 
\label{me3}
\end{eqnarray} 
modulo small off-minus-shell terms. The first term, $-M_0 k^2$, in the brackets 
gives a $G_3$ contribution in Eq.~\ref{me3}, the second and fifth give $G_5$ 
and $G_8$, respectively. Altogether we obtain from Eq.~\ref{me3}
\begin{eqnarray}\nonumber
v_5(12)\psi_N &=& -k^2 M_0
\bar u'_1 \gamma _{\mu } C \bar u'^{T}_{2}\ 
\bar u'_3\gamma ^{\mu }\gamma _5 u_N
-k^2 \bar u'_1 \gamma \cdot P C \bar u'^{T}_{2} \
  \bar u'_3\gamma _5 u_N\\
\nonumber
&& -k^2 \bar u'_1 i\sigma _{\mu \nu }P^{\nu }C\bar u'^{T}_{2}\
\bar u'_3 \gamma ^{\mu }\gamma _5 u_N
+2k\cdot P\bar u'_1\gamma \cdot k C\bar u'^{T}_{2}\
\bar u'_3\gamma _5 u_N\\ \nonumber
&& -2k\cdot P\bar u'_1 i \sigma _{\mu \nu }k^{\nu }C\bar u'^{T}_{2}\
\bar u'_3 \gamma _5 \gamma ^{\mu } u_N
-2M_0\bar u'_1 \gamma \cdot k C \bar u'^{T}_{2} \
\bar u'_3\gamma _5\gamma \cdot k u_N\\ 
&& +2\bar u'_1 i\sigma _{\mu \nu }P^{\mu }k^{\nu }C\bar u'^{T}_{2}\
\bar u'_3\gamma _5\gamma \cdot k u_N.
\label{me4}
\end{eqnarray}      
In Eq.~\ref{me4} the first three terms can be combined to the single 
{\em proper} vector-spin invariant of the nucleon with projector $[P]$ from 
the Melosh boost for each quark, viz.  
\begin{equation}
-{k^2\over 4 M^{2}_{0}}\ \bar u'_1 [P]\gamma _{\mu } C (\bar u'_{2}[P])^{T} 
\;\bar u'_3 [P]\gamma ^{\mu }\gamma _5 u_N,  
\label{aids0}
\end{equation} 
which has only positive energy components. 
The last four terms containing $k$ dependent spin invariants will be
addressed in the next section, where it is shown that they give rise
to N and N*-type spin invariants through the angular $\mbf{k}_{\perp}$
integration. Symbolically we may write Eq.~\ref{me4} as
\begin{equation}
v_5(12)(G_2+G_6) \sim G_3 - G_5 -G_8+\cdots  
\label{me5}
\end{equation} 
in terms of the spin invariants of Table~\ref{tab:1}. This says that a
pseudoscalar interaction $v_5(12)$ between quarks 1 and 2 acting on
the nucleon spin invariants $G_2$ and $G_6$ is off-diagonal and makes
transitions to the nucleon spin invariants $G_3$, $G_5$ and $G_8$ so
far, but more will follow.

We also notice from missing $[P]$'s next to the quark spinors in the $k$ 
dependent terms that small Dirac components are generated by these 
interactions, so that the {\em Pauli-Melosh basis is clearly too limited}. 

The isospin matrix element corresponding to the neutral pion exchange 
$v_5(12)$ is straightforward to evaluate
\begin{eqnarray}
(\chi^{T}_{3}\tau _3 i \tau _2\chi _1)(\chi^{T}_{2}\tau _3\chi _{\uparrow})=1,\ 
\chi _1=|u\rangle,\ \chi _2=|u\rangle,\ \chi _3=|d\rangle. 
\end{eqnarray}
The same result is valid for $1\leftrightarrow 2$. The isospin matrix element 
for the charged pion exchange $v^{+}_{5}(13)$ that we consider next is 
evaluated similarly, viz. 
\begin{eqnarray}
-\chi ^{T}_{3}\tau ^- i\tau _2(\tau ^+)^T \chi _1 (\chi ^{T}_{2} 
\chi _{\uparrow})=-2,
\end{eqnarray}  
where $\tau ^{\pm}=\mp {1\over \sqrt{2}}
(\tau _1\pm i\tau _2),\ \vec\tau \cdot 
\vec\tau=\tau _3\tau _3-(\tau ^+ \tau ^- +\tau ^- \tau ^+).$ 

To demonstrate the case for the $\pi ^+$ exchange interaction we
consider channels 1 and 2, since in the $uds$ basis and for the proton
quarks 1 and 2 are both up quarks and no charged exchange is possible
between them. We consider the 1-2 symmetric combination ${\cal G}_0
[v^{+}_{5}(23)+v^{+}_{5}(32)+v^{+}_{5}(13)+v^{+}_{5} (31)]$ on $G_2$,
where the Green function symbol ${\cal G}_0$ denotes various energy
denominators. To preserve the $uud$ flavor order in the nucleon wave
function we need to relabel the $u$ and $d$ quarks $2\leftrightarrow
3$ and $1\leftrightarrow 3$, respectively. Thus we start from
\begin{eqnarray}\nonumber
\lefteqn{\left[(E^{-1}_{23}
+E^{-1}_{32})v^{+}_{5}(23)+(E^{-1}_{13}+E^{-1}_{31})
v^{+}_{5}(13)\right]{\psi}_N}\\ \nonumber 
&=& 4m^2_q(E^{-1}_{23}+E^{-1}_{32})\ \bar u'_2 \gamma _5 u_2
\ \bar u'_3 \gamma _5 u_3 
\left(\bar u_3 [P]\gamma _5 C\bar u^{T}_{1}\ \bar u_2 u_N+\bar u_3 [P] 
\gamma _5 C \bar u^{T}_{2}\ \bar u_1 u_N\right)\\ \nonumber
&&+4m^2_q(E^{-1}_{13}+E^{-1}_{31})\;\bar u'_1 \gamma _5 u_1\
\bar u'_3 \gamma _5 u_3
\left(\bar u_3 [P] \gamma _5 C\bar u^{T}_{1}\ \bar u_2 u_N
+\bar u_3 [P]\gamma _5 C \bar u^{T}_{2}\ \bar u_1 u_N\right)\\ \nonumber   
&=& -(E^{-1}_{23}+E^{-1}_{32})\ \left[\bar u'_3\gamma _5 \gamma \cdot k [P]C 
\bar u'^{T}_{1}\ \bar u'_2 \gamma \cdot k\gamma _5 u_N\right.\\ \nonumber
\quad &&\left. \qquad\qquad\qquad+\bar u'_3 \gamma _5 \gamma \cdot k [P]\gamma 
\cdot k C\bar u'^{T}_{2}\ \bar u'_1 u_N\right]|udu\rangle
+(1 \leftrightarrow 2)|duu\rangle\\
\nonumber 
&=&-(E^{-1}_{23}+E^{-1}_{32})\ \left[\bar u'_2 \gamma _5\gamma \cdot k[P]
\gamma _5 C \bar u'^{T}_{1}\ \bar u'_3 \gamma \cdot k\gamma _5 u_N\right.\\ 
&&\left.\qquad\qquad\qquad+\bar u'_2\gamma _5 \gamma \cdot k[P]
\gamma \cdot kC\bar u'^{T}_{3}\ \bar u'_1 
u_N\right] |uud\rangle+(1\leftrightarrow 2)\ |uud\rangle
\label{me6}
\end{eqnarray}
requiring no Fierz rearrangement for the (12)3 and (21)3 terms. The 
flavor dependence is displayed on the right hand side of Eq.~\ref{me6} to 
avoid confusion in a comparison with Eq.~\ref{me1}. Equation~\ref{me6} will 
be continued in the next Section after the angular $\mbf{k}_{\perp}$ 
integration.   

\section{Relativistic Coupled Equations}

We now turn to evaluate the integral given in Eq.~\ref{eqn:iter} in
first order approximation.  The angular $\mbf{k}_{\perp}$ integration
is performed using the identity
\begin{equation}
\int d^2\mbf{k}_{\perp} \mbf{a}_{\perp}\cdot \mbf{k}_{\perp}\  
\mbf{b}_{\perp}\cdot 
\mbf{k}_{\perp} f(k^{2})=\pi \mbf{a}_{\perp}\cdot \mbf{b}_{\perp}\int 
dk^{2}_{\perp} {k}^{2}_{\perp}f(k^{2}).  
\label{aid1}
\end{equation}  
If the constant vectors $a$ and $b$ are the Dirac gamma matrices we write for 
two four-vector products  
\begin{equation}
\int d^2\mbf{k}_{\perp} (\gamma )_1\cdot k \ (\gamma )_2\cdot k f(k^{2})
=\pi (\gamma )_1\cdot (\gamma )_2\int dk^{2}_{\perp} k^2 f(k^{2}).  
\label{aid2}
\end{equation}  
modulo $\gamma ^{\pm}$ terms that we ignore at first. Here the sub-indices 
stand for the first or second matrix element in nucleon wave function 
components where the $\gamma $'s occur.

Using 
\begin{equation}
\int d^2\mbf{k}_{\perp} k\cdot P \gamma \cdot k f(k^2) = \gamma \cdot P\ 
\pi \int dk^{2}_{\perp} k^2 f(k^2), 
\label{aid3}
\end{equation}  
the remaining spin invariants in Eq.~\ref{me4} are now approximated by  
\begin{eqnarray}\nonumber
\int d^2\mbf{k}_{\perp} k\cdot P\ \bar u'_1\gamma \cdot k\ C\bar u^{T}_{2}\ 
f(k^2) &=& \bar u'_1\gamma \cdot P C\bar u^{T}_{2}\ \pi \int dk^{2}_{\perp} k^2 
f(k^2),\\ \nonumber
\int d^2\mbf{k}_{\perp} k\cdot P\ \bar u'_1 i \sigma _{\mu \nu }k^{\nu } C
\bar u^{T}_{2}\; f(k^2)&=& \bar u'_1 i \sigma _{\mu \nu }P^{\nu }C\bar u^{T}_{2}
\ \pi \int dk^{2}_{\perp} k^2 f(k^2),\\ \nonumber
\int d^2\mbf{k}_{\perp}\bar u'_1\gamma \cdot k C\bar u^{T}_{2}\ \bar u_3\gamma 
_5\gamma \cdot ku_N\ f(k^2)&=&\bar u'_1\gamma _{\mu } C\bar u^{T}_{2} \ \bar 
u_3\gamma _5\gamma ^{\mu }u_N\ \pi \int dk^{2}_{\perp} k^2 f(k^2),\\
\int d^2\mbf{k}_{\perp}\bar u'_1 i 
\sigma _{\mu \nu }P^{\mu }k^{\nu }C\bar u^{T}_{2}\ 
\bar u_3\gamma _5\gamma \cdot ku_N\;f(k^2)&=&\bar u'_1 i \sigma _{\mu \nu }
P^{\mu }C\bar u^{T}_{2}\ \bar u_3\gamma _5\gamma ^{\nu }u_N\; \pi \int 
dk^{2}_{\perp} k^2 f(k^2). 
\label{me8}
\end{eqnarray} 

At this stage we have to specify $f(k^2)$ for each case in more
detail.  For $v_5(12)$ the kinematics $p'_1=p_1-k$, $p'_2=p_2+k$
mentioned above imply that $E_{12}$ (see Eq.~\ref{eqn:den}) and
$R_0(q^{2}_{3},Q^{2}_{3})$ will depend on $\mbf{k}_{\perp}$ via
$\mbf{p}'_{1\perp}\cdot \mbf{k}_{\perp}$, $\mbf{p}'_{2\perp}\cdot
\mbf{k}_{\perp}$, and $\mbf{q}'_{3\perp}\cdot \mbf{k}_{\perp}$ where
the final momenta (with $'$) are held fixed. In fact,
$(E_{12})^{-1}+(E_{21})^{-1}$, $R_0$ and $M_0$ depend on $q'_3\cdot
k$, as required by translation invariance. Note that $E_{12}$ contains
$-\mbf{k}_{\perp}\cdot \mbf{p}'_{2\perp}$ via $E_2$, while $E_{21}$
contains $+\mbf{k}_{\perp}\cdot \mbf{p}'_{1\perp}$ via $E_1$,
etc. which combine to $\mbf{q}'_3\cdot \mbf{k}_{\perp}=-q'_3\cdot k$.

Therefore, we may display the angular $\mbf{k}_{\perp}$ dependence by 
expanding  
\begin{eqnarray}
R_0(q^{2}_{3},Q^{2}_{3})=(R_0)_0+q'_3\cdot k (R_0)_{q_3}
+Q'_3\cdot k (R_0)_{Q_3}+\cdots,
\label{exp1}
\end{eqnarray}   
where $(R_0)_0$, $(R_0)_{q_3}$ etc. depend on the relative momentum variables 
$q'^{2}_{3}, Q'^{2}_{3}$ and on $k^{2}_{\perp}$ only without angular 
$k_{\perp}$ dependence. Similar expansions are 
valid for suitable combinations of energy denominators and for 
$M_0(q^{2}_{3},Q^{2}_{3})$.  
    
Now the last four terms in Eq.~\ref{me4} in conjunction with Eq.~\ref{me8} 
after the angular integration may be written as 
\begin{eqnarray}
\nonumber
\lefteqn{\int d^2\mbf{k}_{\perp} k\cdot P \bar u'_1 
\gamma \cdot k C\bar u'^{T}_{2} \
\bar u'_3\gamma _5 u_N\,R_0\left({1\over E_{12}}+
{1\over E_{21}}\right)}\\ 
&=&\bar u'_1 \gamma \cdot P C\bar u'^{T}_{2}\ \bar u'_3
\gamma _5 u_N\, \pi \int dk^{2}_{\perp} k^2 (R_0)_0
\left({1\over E_{12}}+{1\over E_{21}}\right)_0 +\cdots,
\label{me9}
\\\nonumber 
\lefteqn{\int d^2\mbf{k}_{\perp} k\cdot P \bar u'_1 
i\sigma _{\mu \nu }k^{\nu } C
\bar u'^{T}_{2}\ \bar u'_3 \gamma _5 \gamma ^{\mu } u_N\, R_0\left({1\over 
E_{12}}+{1\over E_{21}}\right)}\\ 
&=&\bar u'_1 i\sigma _{\mu \nu }P^{\nu }C\bar u'^{T}_{2}\
\bar u'_3 \gamma _5 \gamma^{\mu } u_N\, \pi \int dk^{2}_{\perp} k^2 
(R_0)_0\left({1\over E_{12}}+{1\over E_{21}}\right)_0+\cdots , 
\label{me10}
\\\nonumber 
\lefteqn{\int d^2\mbf{k}_{\perp}
\bar u'_1 \gamma \cdot k C \bar u'^{T}_{2}\ \bar u'_3\gamma _5\gamma \cdot k 
u_N R_0\left({1\over E_{12}}+{1\over E_{21}}\right)}\\ 
&=&\bar u'_1 \gamma _{\mu } C \bar u'^{T}_{2}\;\bar u'_3\gamma _5\gamma ^{\mu } 
u_N\, \pi \int dk^{2}_{\perp} k^2 (R_0)_0\left({1\over E_{12}}+{1\over E_{21}}
\right)_0+\cdots \label{me11}
\\\nonumber 
\lefteqn{\int d^2\mbf{k}_{\perp}\bar u'_1 
i\sigma _{\mu \nu }P^{\mu }k^{\nu }C\bar u'^{T}_{2}\
\bar u'_3\gamma _5\gamma \cdot k u_N\,R_0
\left({1\over E_{12}}+{1\over E_{21}}
\right)}\\ 
&=&\bar u'_1 i\sigma _{\mu \nu }P^{\mu }C\bar u'^{T}_{2}\
\bar u'_3\gamma _5\gamma ^{\nu } u_N\,
 \pi \int dk^{2}_{\perp} k^2 (R_0)_0
\left({1\over E_{12}}+{1\over E_{21}}\right)_0+\cdots.  
\label{me12}
\end{eqnarray}

Before we comment on Eqs.~\ref{me9} to \ref{me12} let us continue the
$\pi ^+$ exchange started with Eq.~\ref{me6}.  Integrating
Eq.~\ref{me6} yields
\begin{eqnarray}
\nonumber
{\cal I}_2&\equiv& \int d^2\mbf{k}_{\perp} \left[v^{+}_{5}(23)\left({1\over 
E_{23}}+{1\over E_{32}}\right)+v^{+}_{5}(13)\left({1\over E_{13}}+{1\over 
E_{31}}\right)\right]R_0\psi_N \\ \nonumber 
&=&-\bar u'_2 \gamma _5\gamma _{\mu }[P]\gamma _5 C \bar u'^{T}_{1}\
\bar u'_3 
\gamma ^{\mu }\gamma _5 u_N\,\pi \int dk^{2}_{\perp} k^2 (R_0)_0 \left({1\over 
E_{23}}+{1\over E_{32}}\right)_0\\ \nonumber
&&-\bar u'_1\gamma _5 \gamma _{\mu }[P]\gamma _5C\bar u'^{T}_{2}\ \bar u'_3 
\gamma ^{\mu }\gamma _5 u_N\,\pi \int dk^{2}_{\perp} 
k^2 (R_0)_0 \left({1\over E_{13}}+{1\over E_{31}}\right)_0\\ \nonumber  
&&-\bar u'_2 \gamma _5 \gamma _{\lambda }[P]\gamma ^{\lambda }C \bar u'^{T}_{3}
\ \bar u'_1 u_N\,\pi \int dk^{2}_{\perp} 
k^2 (R_0)_0 \left({1\over E_{23}}+{1\over E_{32}}\right)_0\\ 
&&-\bar u'_1 \gamma _5\gamma _{\lambda }[P]\gamma ^{\lambda } C \bar u'^{T}_{3}
\ \bar u'_2 u_N\,\pi \int dk^{2}_{\perp} 
k^2 (R_0)_0 \left({1\over E_{13}}+{1\over E_{31}}\right)_0+\cdots. 
\label{me13}
\end{eqnarray}
In Eq.~\ref{me13} the first two terms can be simplified and the last two 
Fierz rearranged to the canonical (12)3 order, so that Eq.~\ref{me13} may be 
symbolically written as 
\begin{eqnarray}\nonumber 
\int d^2\mbf{k}_{\perp} {\cal G}_0\left(v^{+}_{5}(23)+v^{+}_{5}(13)\right)
(G_2+G_6)\sim {3\over 2}G_1\ r_{-}+{3\over 2}G_3\ r_{+}\\ 
+G_6\ R_{-}+{1\over 4}G_7\ r_{+}+2 G_8\ R_{+}+\cdots, \label{me14}
\end{eqnarray}
where $r_{\pm}, R_{\pm}$ are radial integrals defined by 
\begin{equation} 
r_{\pm}=\pi \int dk^{2}_{\perp} k^2 (M_0 R_0)_0 \left[\left({1\over E_{23}}
+{1\over E_{32}}\right)_0\pm \left({1\over E_{13}}
+{1\over E_{31}}\right)_0\right], 
\label{rad1}
\end{equation}
\begin{equation}   
R_{\pm}=\pi \int dk^{2}_{\perp} k^2 (R_0)_0 \left[\left({1\over E_{23}}
+{1\over E_{32}}\right)_0\pm \left({1\over E_{13}}
+{1\over E_{31}}\right)_0\right].  
\label{rad2}
\end{equation}
Clearly Eqs.~\ref{me9} to \ref{me12} separate the last four terms of 
Eq.~\ref{me4} into the nucleon spin invariants $G_3, G_5, G_8$ and radial 
integrals. The trem with $v^{+}_{5}$ now is seen to contain many other  
proton spin invariants other than $G_2$. The ellipses indicate 
new N*-type spin invariants generated from higher order terms in expansions 
like Eq.~\ref{exp1}, which we address next.  

When terms linear in $k\cdot q'_3$ and $k\cdot Q'_3$ are considered as 
exhibited in Eq.~\ref{exp1}, then the relevant replacement of the identity 
Eq.~\ref{aid3} is 
\begin{equation}
\int d^2\mbf{k}_{\perp} k\cdot q'_3 \gamma \cdot k f(k^2) = \gamma \cdot q'_3\ \pi 
\int dk^{2}_{\perp} k^2 f(k^2),  
\label{aid4}
\end{equation}  
etc. This entails replacing the two $\gamma \cdot k$ factors in expressions 
like Eq.~\ref{me1} and \ref{me6} resulting from Eq.~\ref{eqn:iter} by 
$\gamma \cdot q'_3$'s or $\gamma \cdot q'_3 \ \gamma \cdot Q'_3$, etc. in the 
coupled radial equations of motion. The resulting spin invariants are 
characteristic of those of N*'s~\cite{BKW} with one or two quarks in higher 
orbitals. Thus, N* spin invariants in coupled equations for the nucleon 
originate from the entanglement of momentum wave functions (i. e. orbital 
angular momentum) with the 3-quark spin structure.   
 
The first higher order terms contributing to Eq.~\ref{me9}, for example, are 
given by  
\begin{eqnarray}
\nonumber 
&&q'_3\cdot Q'_3\bar u'_1 \gamma \cdot PC\bar u'^{T}_{2} \ \bar u'_3 \gamma _5
u_N\, \pi \int dk^{2}_{\perp} k^4 (R_0)_{Q_3}
\left({1\over E_{12}}+{1\over E_{21}}\right)_{q_3}\\ \nonumber
&&+Q'_{3}\cdot P\bar u'_1 \gamma \cdot q'_3 C\bar u'^{T}_{2}\
\bar u'_3 \gamma _5u_N\, \pi \int dk^{2}_{\perp} k^4 (R_0)_{Q_3}
\left({1\over E_{12}}+{1\over E_{21}}\right)_{q_3}\\ 
&&+q'_{3}\cdot P\bar u'_1 \gamma \cdot Q'_3 C\bar u'^{T}_{2}\
\bar u'_3 \gamma _5u_N\, \pi \int dk^{2}_{\perp} k^4 (R_0)_{Q_3}
\left({1\over E_{12}}+{1\over E_{21}}\right)_{q_3}+\cdots, 
\label{me16}
\end{eqnarray}
where the second invariant vanishes when the Dirac equation is applied and the 
last invariant corresponds to a component of the negative parity 
N*(1535) configuration $S_{11}$. However, upon integrating there are higher 
order continuations for every term in Eq.~\ref{me4}. E. g. the second term 
continues with  
\begin{eqnarray}\nonumber
\int d^2\mbf{k}_{\perp}k^2 R_0 \left({1\over E_{12}}+{1\over E_{21}}\right)
&=&\cdots+q'_3\cdot Q'_3\ \pi \int dk^{2}_{\perp} k^4 (R_0)_{Q_3}
\left({1\over E_{12}}+{1\over E_{21}}\right)_{q_3}\\
&&+q'^{2}_{3}\ \pi \int dk^{2}_{\perp} k^4 (R_0)_{q_3}
\left({1\over E_{12}}+{1\over E_{21}}\right)_{q_3}+\cdots. 
\label{me17}
\end{eqnarray}  
Clearly, these terms represent higher order contributions from the expansion 
of the momentum dependence of the radial wave function and energy denominators. 
\par
Next we turn to the gluon exchange as an example of a vector 
interaction. Again we start from the 1-2 symmetric $v_G(12)$ for simplicity,  
\begin{eqnarray} 
\nonumber 
v_G(12)\psi_N&\equiv&4m_q^2\, \bar u'_1\gamma ^{\mu }u_1\
\bar u'_2\gamma _{\mu }u_2\,\left(
\bar u_3[P]\gamma _5C\bar u^{T}_{2}\;\bar u_1 u_N+\bar u_3[P]\gamma _5 C\bar 
u^{T}_{1}\;\bar u_2 u_N\right)\\ \nonumber 
&=&-\bar u_3[P]\gamma \cdot k\gamma _{\mu }\gamma _5 C\bar u'^{T}_{2}\
\bar u'_1 \gamma ^{\mu }\gamma \cdot k u_N
+4\bar u_3[P]\gamma _5 C\bar u^{T}_{2}\
\bar u'_1 u_N\,p'_1\cdot p'_2+(1\leftrightarrow 2)\\
&&+2\bar u_3 [P] \gamma _5 C\bar u'^{T}_{2}\;\bar u'_1\gamma \cdot p'_2\gamma 
\cdot k u_N-2\bar u_3 [P] \gamma \cdot k \gamma \cdot p'_1 \gamma _5 C
\bar u'^{T}_{2} \ \bar u'_1 u_N-(1\leftrightarrow 2). 
\label{g1}
\end{eqnarray}
By Fierz rearrangement Eq.~\ref{g1} leads to the following results with  
canonical (12)-spin matrix elements   
\begin{eqnarray}
\nonumber 
v_G(12)\psi_N&=&-\bar u'_1\gamma _{\lambda }C\bar u'^{T}_{2}\ 
\bar u_3[P]\gamma 
\cdot k \gamma _5\gamma ^{\lambda }\gamma \cdot k u_N+2p'_1\cdot p'_2
\left[2 M_0 
\bar u'_1 \gamma _{\lambda }C\bar u'^{T}_{2}\; \bar u_3\gamma _5\gamma ^{
\lambda }u_N\right.  \\ \nonumber 
&&\left. -2\bar u'_1\gamma \cdot P C\bar u'^{T}_{2}\
 \bar u_3\gamma _5 u_N+\bar u'_1 
i\sigma _{\mu \nu }P^{\nu }C\bar u'^{T}_2\
\bar u_3\gamma _5\gamma ^{\mu }u_N\right]
\\ \nonumber
&&+{1\over 2}\,\bar u'_1 C \bar u'^{T}_{2}\
\bar u_3[P]\Big[-\gamma \cdot k \gamma 
\cdot (p'_1+p'_2)\gamma _5+\gamma _5 
\gamma \cdot (p'_1+p'_2)\gamma \cdot k\Big]
u_N\\ \nonumber 
&&+{1\over 2}\,\bar u'_1 \gamma _{\mu }C \bar u'^{T}_{2}\
\bar u_3[P]\Big[-\gamma 
\cdot k \gamma \cdot (p'_1-p'_2)\gamma _5\gamma ^{\mu }-\gamma _5 
\gamma ^{\mu }\gamma \cdot (p'_1-p'_2)\gamma \cdot k\Big]u_N\\ \nonumber 
&&+{1\over 4}\,\bar u'_1 \sigma _{\mu \nu }C \bar u'^{T}_{2}\
\bar u_3[P]\Big[-\gamma 
\cdot k \gamma \cdot (p'_1-p'_2)\gamma _5\sigma ^{\mu \nu }-\gamma _5 
\sigma ^{\mu \nu }\gamma \cdot (p'_1-p'_2)\gamma \cdot k\Big]u_N\\ \nonumber  
&&+{1\over 2}\,\bar u'_1 \gamma _5\gamma _{\mu }C \bar u'^{T}_{2}\
\bar u_3[P]
\Big[\gamma \cdot k \gamma \cdot (p'_1+p'_2)\gamma ^{\mu }-\gamma ^{\mu }\gamma 
\cdot (p'_1+p'_2)\gamma \cdot k\Big]u_N\\ 
&&+{1\over 2}\,\bar u'_1 \gamma _5C \bar u'^{T}_{2}\
\bar u_3[P]\Big[-\gamma \cdot k 
\gamma \cdot (p'_1+p'_2)+\gamma \cdot (p'_1-p'_2)\gamma \cdot k\Big]u_N,
\label{g2}
\end{eqnarray}
which contain terms linear in $k$ and the $p'_i$. As was the case for
the pion exchange, for gluon exchange the explicit $\gamma \cdot p'_i$
factors lead to N*-type spin invariants here also. In the
$d^2\mbf{k}_\perp$-integration, a la Eq.~\ref{aid2}, the linear term
$\gamma \cdot k$ combines with $q'_3\cdot k$ from the $R_0$ expansion
Eq.~\ref{exp1}, when $\gamma \cdot (p'_1-p'_2)$ is present as well in
Eq.~\ref{g2}, or with $Q'_3\cdot k$ when $\gamma \cdot (p'_1+p'_2)$ is
present, etc.\footnote{Strictly speaking, a four-vector p must be
replaced by $\tilde p=p-P p\cdot P/P^2$ to maintain orthogonality. We
omit the tilde for simplicity.}  Thus, Eq.~\ref{g2} leads to the
following spin matrix elements which can be readily manipulated into
canonical form:
\begin{eqnarray*}
\bar u'_1\gamma _{\lambda }C\bar u'^{T}_{2}\ \bar u_3[P]\gamma _{\mu } 
\gamma _5\gamma ^{\lambda }\gamma ^{\mu } u_N,
\end{eqnarray*}
for the first term, i. e. the vector invariant $G_3$ of
Table~\ref{tab:1}, while the third term gives $G_5$ and the fourth
$G_8$. The second term in Eq.~\ref{g2} also gives $G_3$. For the
scalar (12)-term $\bar u'_1 C \bar u'^{T}_{2}$ the associated second
spin matrix elements reduce to
\begin{eqnarray}
-\bar u_3[P]\gamma \cdot Q'_3 \gamma \cdot (p'_1+p'_2)\gamma _5 u_N+
\bar u_3[P]\gamma _5\gamma \cdot (p'_1+p'_2)\gamma \cdot Q'_3 u_N
=-4M^{2}_{0}(1-x'_3)\bar u_3\gamma _5\gamma \cdot Q'_3u_N,  
\label{g3}
\end{eqnarray}
i.e. an N*-spin invariant.  For the vector (12)-term $\bar u'_1 \gamma
_{\mu }C \bar u'^{T}_{2}$ they are
\begin{eqnarray}
\lefteqn{-\bar u'_3[P]\gamma \cdot q'_3 \gamma \cdot 
(p'_1-p'_2)\gamma _5\gamma ^{\mu }
u_N- \bar u'_3[P]\gamma _5\gamma ^{\mu }\gamma 
\cdot (p'_1-p'_2)\gamma \cdot 
q'_3u_N}\nonumber\\&=&[-4P^{\mu }q'_3\cdot (p'_1-p'_2)
+4q_3^{\prime\mu }P\cdot (p'_1-p'_2)]\bar 
u'_3\gamma _5u_N\nonumber\\
&&+4M_0 q'_3\cdot (p'_1-p'_2)\bar u'_3\gamma _5\gamma ^{\mu }u_N
-4M_0q_3^{\prime\mu }\bar u'_3\gamma _5\gamma \cdot (p'_1-p'_2)u_N
\label{g4}   
\end{eqnarray}
i. e. $G_5, G_1, G_3$ N-spin and an N*-spin invariant. Terms with
$\gamma \cdot q'_3$ replaced by $\gamma \cdot Q'_3$ are similarly
reduced to the canonical forms of Table 1. The tensor (12)-term of
Eq.~\ref{g3} generates $G_7$ and different N*-spin invariants, the
axialvector (12)-term generates $G_6$ and N*-spin invariants, while
the pseudoscalar (12)-term generates $G_2$ and three different N*-spin
invariants. In summary the factorization into radial integrals and spin 
invariants for the gluon exchange may be written symbolically as follows 
\begin{eqnarray}
\int d^2\mbf{k}_{\perp}v_G(12)\psi _N
&=&\pi \int dk^{2}_{\perp} (R_0)_0
\Big({1\over E_{12}}+{1\over E_{21}}\Big)_0
\Big[4 M_0 (k^2+p'_1\cdot p'_2) G_3\nonumber\\&&\qquad\qquad-4 
(k^2+p'_1\cdot p'_2) G_5 +2i\ p'_1\cdot p'_2 G_7\Big]\nonumber\\
&&+(R_0)_{Q_3}\Big({1\over E_{12}}+{1\over E_{21}}\Big)_0\Big[
-2M^{2}_{0}(1-x'_3)\bar u'_{1}\gamma \cdot 
PC\bar u^{'T}_{2}\;\bar u'_3\gamma _5
\gamma \cdot Q'_3 u_N\Big] \nonumber\\
&&+(R_0)_{q_3}\Big({1\over E_{12}}
+{1\over E_{21}}\Big)_0 \Big[-2q'_3\cdot
(p'_1-p'_2)G_5+ 2 {(x'_2-x'_1)m_q\over (1-x'_3)}
 P\cdot (p'_1-p'_2)G_1
\nonumber\\ &&+2M_0q'_3\cdot (p'_1-p'_2)
G_3-2M_0{(x'_2-x'_1)m_q
\over 1-x'_3}\bar u'_1C\bar u^{'T}_{2}\; \bar u'_3\gamma _5\gamma \cdot 
(p'_1-p'_2)u_N\Big]\nonumber\\
&&+(R_0)_{Q_3}\Big({1\over E_{12}}+{1\over E_{21}}\Big)_0
\Big[-2M_0\bar u'_1\gamma _5
 C\bar u^{'T}_2\; \bar u'_3\gamma \cdot 
Q'_3 \gamma \cdot (p'_1+p'_2)u_N
\nonumber\\
&&+M_0Q'_3\cdot (p'_1+p'_2)\bar u'_1\gamma _5 C
\bar u^{'T}_2\; \bar u'_3u_N
-2Q'_3\cdot P\bar u'_1\gamma _5 C\bar u^{'T}_2\; \bar u'_3
\gamma \cdot (p'_1+p'_2)u_N\nonumber\\
&&+2P\cdot (p'_2+p'_2)\bar u'_1\gamma _5 C\bar 
u^{'T}_2\; \bar u'_3 \gamma \cdot Q'_3u_N\Big]+\dots
\label{g5}
\end{eqnarray}
 
\section{Conclusions}  

The light front form provides a framework to treat few-body systems in a
relativistic covariant way. This approach is particularly suited when 
boosts are important, which is the case, e.g., for form factor
calculations. So far nucleon light front wave functions have been
treated in a restricted way in this context, namely using positive
energy projected wave function components only that can be directly
derived from the Pauli-Melosh basis. Besides this
restriction, the spatial part of the wave function have been
assumed symmetric and in the ``ground state''. In this paper we have
addressed this issue and shown that residual interactions between the
quarks lead to new components in the wave functions. These can be
compared to configuration mixing in the nonrelativistic approach and
have not been studied so far. They are expected to play an important
role for the mass splittings, the $E2/M1$ ratio of the $\Delta $N
transition, the different high energy behavior of magnetic and
electric form factors of the nucleon, and other important structure
information of the baryons. For example, since small Dirac components are 
included in the spin-flavor invariants, relativistic effects such as some pair 
currents are already included in the impulse approximation. (Compare Refs.
~\cite{bon98,car98} for the two-nucleon case.) We
have investigated two currently used residual interactions,
i.e. $\gamma _5$ coupling representing a typical Goldstone boson
exchange and $\gamma _\mu $ coupling representing the gluon exchange and
found many new spin-isospin structures not present in the simple
nucleon wave function.

The $uds$ basis used in this context suggests working in the Faddeev
formalism for the three body system. We have evaluated a few of the
integrals appearing in the respective equations utilizing light front
time ordered perturbation theory. As a full solution of the problem
still needs further work a next step in this direction is a one
iteration approximation that has been proven quite useful in the two
nucleon case and might be useful in the quark context also, because of
the perturbative nature of the Goldstone boson exchange (that is based
on chiral perturbation theory).

\section{Acknowledgements} The authors gratefully acknowledge the
warm hospitality extended to them during stays at each others home
institutions. 

\begin{table}
\caption{\label{tab:1} Relativistic spin invariants for the
nucleon. All invariants are symmetric with respect to exchange of
12. With respect to spin or isospin alone $G_2$, $G_4$, and $G_6$ are
antisymmetric in the pair 12, the others are symmetric.}
\[
\begin{array}{lrcl}\hline\hline
G_1:& 1G&\otimes &\gamma _{5}u_{\lambda }\\
G_2:& \gamma _{5}G&\otimes &u_{\lambda }\\
G_3:& \gamma ^{\mu }\mbf{\tau }G&\otimes &\gamma _{5}\gamma _{\mu }
\mbf{\tau }u_{\lambda }\\
G_4:& \gamma ^{\mu }\gamma _{5}G&\otimes& \gamma _{\mu }u_{\lambda }\\
G_5:& \gamma \cdot P\mbf{\tau }G&\otimes&\gamma _{5}\mbf{\tau }u_{\lambda }\\
G_6:& \gamma \cdot P\gamma _{5}G&\otimes &u_{\lambda }\\
G_7:& \sigma ^{\mu \nu }\mbf{\tau }G&\otimes
&\gamma _{5}\sigma _{\mu \nu }\mbf{\tau }u_{\lambda }\\
G_8:& i\sigma ^{\mu \nu }P_{\nu }\mbf{\tau }G
&\otimes &\gamma _{5}\gamma _{\mu }
\mbf{\tau }u_{\lambda }\\\hline\hline
\end{array}
\]
\end{table}

\begin{table}
\caption{\label{tab:2} Fierz rearrangements for scalar (S), vector
(V), tensor (T), axial vector (A), and pseudoscalar (P) operators
(matrices). For definitions see Ref.~\protect\cite{WAP}.}
\[
\begin{array}{rrrrrr}
\hline\hline 
&S&V&T&A&P\\
\hline
S&\frac{1}{4}&\frac{1}{4}&\frac{1}{8}&-\frac{1}{4}&\frac{1}{4}\\
V&1&-\frac{1}{2}&0&-\frac{1}{2}&-1\\
T&3&0&-\frac{1}{2}&0&3\\
A&-1&-\frac{1}{2}&0&-\frac{1}{2}&1\\
P&\frac{1}{4}&-\frac{1}{4}&\frac{1}{8}&\frac{1}{4}&\frac{1}{4}\\
\hline\hline
\end{array}
\]
\end{table}


\end{document}